\documentclass[%
 reprint,
prapplied,
 amsmath,amssymb,
 aps,
]{revtex4-2}

\usepackage{graphicx}% Include figure files
\usepackage{dcolumn}% Align table columns on decimal point
\usepackage{bm}% bold math
\usepackage{siunitx}
\usepackage{hyperref}% add hypertext capabilities
\usepackage{upgreek}
\begin{document}

\preprint{APS/123-QED}

\title{Optically-trapped microspheres are high-bandwidth acoustic transducers}

\author{L.~E.~Hillberry}
\email{lhillber@utexas.edu}
\author{M.~G.~Raizen}
\affiliation{Department of Physics, The University of Texas at Austin, Austin, Texas 78712, USA}

\date{\today}

\begin{abstract}
We report on the use of an optically-trapped microsphere as an acoustic transducer.
A model for the hydrodynamic coupling between the microsphere and the surrounding acoustic fluid flow is combined with thermo-mechanical calibration of the microsphere's position detection to enable quantitative acoustic measurements. We describe our technique in detail, including the self-noise, sensitivity, and minimum detectable signals, using a model appropriate for both liquid and gas environments. We then test our approach in an air-based experiment and compare our measurements with two state-of-the-art commercially-available acoustic sensors. Piezoelectrically-driven bursts of pure tones and laser ablation provide two classes of test sounds. We find accurate measurements with a bandwidth of 1 MHz are possible using our technique, improving by several orders of magnitude the bandwidth of previous flow measurements based on optically-trapped microspheres.
\end{abstract}
\maketitle

%\tableofcontents

\section{Introduction}

Owing to their micro-manipulation and force transduction capabilities, optical tweezers have become an indispensable tool in a variety of scientific fields \cite{volpe_roadmap_2023}. By tightly focusing a laser beam, optical forces can exceed gravitational forces and thermal fluctuations to stably trap micron-scale objects \cite{ashkin_acceleration_1970}. In vacuum \cite{ashkin_optical_1976}, optical tweezers have enabled zeptonewton force sensing~\cite{ranjit_zeptonewton_2016}, state-of-the-art torque sensitivity~\cite{ahn_ultrasensitive_2020}, and searches for new physics~\cite{moore_searching_2021}, including proposals to measure high-frequency gravity waves~\cite{arvanitaki_detecting_2013}. Also in vacuum, optical tweezers can trap and cool microspheres \cite{li_millikelvin_2011} to the motional ground state \cite{delic_cooling_2020, piotrowski_simultaneous_2023}, and have been multiplexed to arrays of hundreds of single-atom traps in a promising platform for quantum computation and simulation \cite{ebadi_quantum_2021, bluvstein_quantum_2022}. In aqueous solution, optical tweezers can measure mechanical properties of life at the nano-scale \cite{ashkin_bacteria_1987,ashkin_cells_1987}, such as the stepping strength of molecular motors or the rigidity of bio-molecules \cite{Gennerich_optical_2017, svoboda_biological_1994,svoboda_direct_1993}. Also in liquid, optical tweezers enable ultra-fast viscosity measurements~\cite{madsen_ultrafast_2021} and Casimir force measurements~\cite{pires_probing_2021}. In gaseous media, optical tweezers have revolutionized single-particle aerosol science \cite{kalume_optical_2021}, including absolute pressure measurements and species identification \cite{blakemore_absolute_2020}, mass metrology \cite{hillberry_weighing_2020, carlse_technique_2020}, and single-droplet growth and freezing studies \cite{ashkin_optical_1975, magome_optical_2003,ishizaka_in_2011}.

\begin{figure}
\includegraphics[width=\columnwidth]{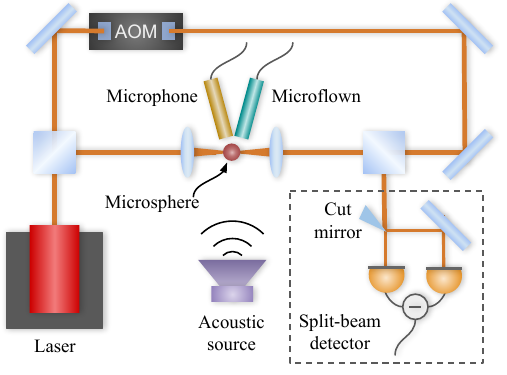}
\caption{Schematic depiction of the experimental set up. A 1064 nm laser is split by a polarizing beamsplitter. The $p$-polarized beam is sent through an acousto-optic modulator (AOM) to shift its frequency by 80 MHz, thereby eliminating interference effects in the trap. The $p$-polarized beam is then steered counter-propagating to the $s$-polarized beam and both are focused to the same point between twin aspheric lenses (numeric aperture 0.7), generating a stable optical trap for silica microspheres in air. After passing through the trap, the $s$-polarized beam is separated with a second polarizing beamsplitter and sent to the detection system. For detection, a sharp, D-shaped cut mirror splits the incoming transverse mode into two halves that are sent to a balanced photo-detector (75 MHz bandwidth).  Various acoustic sources provide test sounds, and additional acoustic sensors, a microphone and Microflown, are positioned just behind the trap. The entire system is enclosed in a multi-chamber acrylic box to mitigate air currents.   \label{fig:schematic} 
    }
\end{figure}

There further exists a body of work using optically-trapped microspheres to measure flow in liquids \cite{kirchner_direct_2014, nedev_microscale_2014, ohlinger_optically_2012,dehnavi_optical_2020, bruot_direct_2021, bassindale_measurements_2014}. So far, these studies have characterized low frequency ($<500$ Hz) flows by monitoring the motion of optically-trapped microspheres with a camera or position-sensitive detector. In this Letter, we propose and demonstrate a fluid velocity measurement scheme with a bandwidth approaching 1 MHz using optically-trapped microspheres in air. Flow at such high frequencies is generally associated with acoustic radiation. A schematic of our optically-trapped-microsphere acoustic transducer is shown in Fig.~\ref{fig:schematic}. Other non-traditional acoustic sensors have recently been studied, including optical micro-resonators \cite{basiri-esfahani_precision_2019, tang_single_2023}, and laser deflection or interference methods \cite{wissmeyer_looking_2018}. As we will see, our method uniquely combines self-calibration, high-bandwidth, and high-sensitivity to acoustic velocity waves (rather than pressure waves).

Our method builds on earlier work that first measured the instantaneous velocity of a thermally-fluctuating microsphere in air \cite{li_measurement_2010}. This same system is not only sensitive to thermal fluctuations, but also to acoustic perturbations. Two ingredients, a hydrodynamic model of the acoustic force and thermo-mechanical self-calibration, enable quantitative acoustic measurements. Since the microsphere is uniquely sensitive to high frequency velocity flows, we use two commercially-available sensors to asses our platform's capabilities: We benchmark our method in terms of accuracy and bandwidth against 1) a high-bandwidth (\qty{200}{kHz}) pressure microphone, and
%whose entire body including the preamplifier, is less than 1/8-inch in diameter, making it the smallest measurement microphone set in the world~\cite{GRAS_worlds_2017}. 
2) a micron-scale dual-hot-wire anemometer \cite{bree_three-dimensional_1999, bree_overview_2003} (calibrated bandwidth \qty{20}{kHz}) that is commercially known as the \emph{Microflown} \cite{bree_microflown_2009}. 

The remainder of this paper is organized as follows: In Section \ref{sec:model} we describe the microsphere's acoustic sensing modality, including calibration, self-noise, and minimum-detectable signals. Section \ref{sec:results} reports our sound detection results. We then discuss our results within the context of other microsphere-based flow measurements and speculate on future applications in Section \ref{sec:discussion}. The paper is then concluded in Section \ref{sec:conclusions}.

\section{Noise, calibration, and acoustic response} \label{sec:model}

In thermal equilibrium with a reservoir fluid at finite temperature, a microsphere's position fluctuates in random and perpetual \emph{Brownian motion} \cite{brown_brief_1828}. Brownian motion velocity detectors \cite{li_measurement_2010,kheifets_high-sensitivity_2014,madsen_ultrafast_2021} are sensitive to both thermally fluctuating and driven fluid flows. If the resulting driven motion is larger than the random thermal motion (and detector noise), an acoustic signal is detectable. In what follows we develop a model for the acoustic signal and thermal noise of our proposed acoustic detection system. 

For the general setup, consider a microsphere of radius $R$ and density $\rho$ harmonically bound to the coordinate origin. The microsphere mass is $m=4 \pi \rho R^3/3$ and the harmonic trap strength is $\kappa$. 
Let the trapping fluid at temperature $T$ have density $\rho_{\rm f}$, speed of sound $c_0$, and dynamic viscosity $\eta$. 
The $x$-component of the system's equation of motion is
\begin{equation} \label{eq:eom}
    m \ddot{x}(t) + \kappa x(t) - F_{\rm d}[v(t)] =  F_{\rm ext}(t) + F_{\rm th}(t)
\end{equation}
where $v(t)=\dot{x}(t)$ is the microsphere's velocity at time $t$, $F_{\rm d}(v)$ is the dissipative, velocity-dependent drag force, and $F_{\rm ext}$ is an external driving force. $F_{\rm th}$ is the fluctuating thermal force that is related to the dissipative force through the fluctuation-dissipation theorem.

When all bounding walls are far from the sphere \cite{mo_broadband_2015} and the fluid flow at sphere's surface does not slip \cite{premlata_atypical_2020}, the hydrodynamic drag force in the incompressible limit is~\cite{basset_motion_1888,maxey_equation_1983,temkin_velocity_1976}
\begin{equation}
    F_{\rm d}[v(t)] = - \gamma_0 \left (v(t) + \sqrt{\frac{\tau_{\rm f}}{\pi}}\int_{-\infty}^{t} {\rm d}t'\, \frac{\dot{v}(t')}{\sqrt{t-t'}} \right ) -\frac{\delta}{2} m \dot{v}(t) \, , \label{eq:basset}
\end{equation}
where $\gamma_0 = 6 \pi \eta R$ is the Stokes friction coefficient and $\delta=\rho_{\rm f} / \rho$ is the fluid-to-microsphere density ratio. 
The \emph{vorticity diffusion time} 
$\tau_{\rm f} = R^2 \rho_{\rm f}/\eta = 9\delta \tau_{\rm p}/2$
is the amount of time it takes for vorticity --- the curl of velocity --- to diffuse across the sphere and $\tau_{\rm p} = m/\gamma_0$  is the momentum diffusion time. The first, second, and third terms of Eq.~\eqref{eq:basset} describe, respectively, Stokes drag (independent of $\delta$), viscous damping due to the flow history (proportional to $\delta^{1/2}$), and inertia of the mass added by the fluid that follows the microsphere (proportional to $\delta$). For a silica microsphere in air $\delta\sim\si{10^{-3}}\ll1$ hence Eq.~\eqref{eq:basset} reduces to $F_{\rm d}[v(t)]\approx-\gamma_0 v(t)$.

In the frequency domain, we may write $F_{\rm d}[v(\omega))]=-\gamma(\omega)v(\omega)$ where $\omega=2 \pi f$ is the circular frequency, the frequency-dependent damping is~\cite{stokes_effect_1851,landau_fluid_2013}
\begin{equation}\label{eq:hydro_damping}
    \gamma(\omega) = \gamma_0 \left (1 + \sqrt{-i \tau_{\rm f} \omega} - i\frac{\tau_{\rm f} \omega}{9} \right )\, ,
\end{equation}
and $\sqrt{-i} = (1-i)/\sqrt{2}$ defines the square-root's branch cut. 

Next, we consider two cases: \emph{noise} when $F_{\rm ext} = 0$ and \emph{signal} when $F_{\rm th} = 0$ and $F_{\rm ext}$ is caused by an acoustic wave.
\begin{figure}[htbp]
\includegraphics[width=\columnwidth]{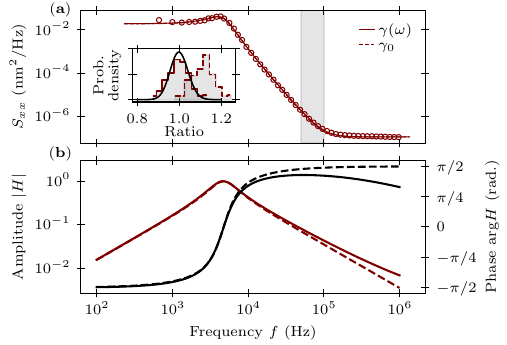} 
\caption{(a) Experimental position power spectral density (open circles) of a $R=\qty{1.51(5)}{\micro m}$ silica microsphere thermally driven by air at $T=\qty{23.97(1)}{\celsius}$ with a relative humidity of 57(1)\%, which has a viscosity $\eta=\qty{18.23(1)}{\micro Pa\, s}$ \cite{tsilingiris_review_2018}. The experimental spectrum is an average periodogram of 550 signals of length 3 ms. For visualization, each point of the experimental spectrum is an average over logarithmically-spaced frequency bins. Calibration is performed by fitting the voltage spectrum in the 1 kHz to 30 kHz band to Eq.~\eqref{eq:voltage_PSD_air} (dashed line). The spectrum and fit are shown here in physical units using the calibration result. The solid line uses the fit results to include hydrodynamic effects that are imperceptible up to $\sim \qty{50}{kHz}$. However, the 50 kHz to 100 kHz band (gray shaded region) does exhibit subtle hydrodynamic effects, as suggested by the the data-to-theory ratio's probability density (inset), wherein the hydrodynamic theory (solid red line) follows much more closely the expected Erlang distribution of ratios (solid black line) \cite{norrelykke_power_2010,hillberry_weighing_2020}. (b) Theoretical transfer function relating microsphere velocities to fluid velocities. The red lines show the amplitude on the left axis while the black lines show the phase on the right axis. The solid line corresponds to the hydrodynamic theory while the dashed lines makes the approximation $\gamma(\omega) \approx \gamma_0.$ The microsphere, trap, and fluid parameters are chosen to be consistent with the calibration shown in (a).
\label{fig:calibration} 
    }
\end{figure}
\subsection{Noise}
The thermal force is $F_{\rm th}(t) = \sqrt{2 k_{\rm B}T \gamma_0}\xi(t)$ where $\xi(t)$ is a zero-mean,
possibly-time-correlated \cite{tothova_note_2016} random variable, and $k_{\rm B}$ is Boltzmann's constant.
When $F_{\rm ext}=0$, the equation of motion \eqref{eq:eom} may be solved in the frequency domain for the \emph{admittance} $v(\omega) / F_{\rm th}(\omega) = (\gamma(\omega) - i \omega m + i \kappa / \omega)^{-1}$
The corresponding (one-sided) velocity power spectral density is given by the Kubo-Green formula ~\cite{kubo_fluctuation-dissipation_1966} as
\begin{equation} \label{eq:velocity_PSD}
    S_{vv}(\omega) = 4 k_{\rm B} T \, {\rm Re} \left [ (\gamma(\omega) - i \omega m + i \kappa / \omega)^{-1} \right ]\,.
\end{equation}
%Hence, if driven by $F_{\rm ext}$, the resulting $v(\omega)$ exceeds thermal fluctuations when $v^2(\omega) > S_{vv}(\omega)$. 
Equation \eqref{eq:velocity_PSD} describes the microsphere's thermal fluctuations and hence the inherent noise which must be overcome to detect $F_{\rm ext}\neq0$. However, beyond noise limitations, thermal fluctuations enable an accurate detector calibration scheme.

The split-beam detection method, depicted in Fig.~\ref{fig:schematic}, generates a linear voltage signal $V(t) = \beta x(t)$ where $\beta$ is the displacement-to-voltage calibration factor. For silica microspheres in air, the radius $R$, temperature $T$, and viscosity $\eta$ can be considered known to within a couple percent \cite{hillberry_weighing_2020,tsilingiris_review_2018,stober_controlled_1968}. Since $S_{xx}=S_{vv}/\omega^2$ we can predict the detector's (one-sided) Brownian-motion-driven voltage power spectral density 
\begin{align}
    S_{VV}(\omega) &= \frac{\beta^2}{\omega^2} S_{vv}(\omega) \label{eq:voltage_PSD}\\ 
    &\approx \beta^2 \frac{4 k_{\rm B} T \gamma_0}{(m \omega ^2 - \kappa)^2 + \gamma_0^2\omega^2} \, . \label{eq:voltage_PSD_air}
\end{align}
The second approximate equality ~\eqref{eq:voltage_PSD_air} is accurate for thermal fluctuations in air and assumes $\gamma(\omega)\approx \gamma_0$.
As shown in Fig.~\ref{fig:calibration} (a), by averaging experimental periodograms of thermally-driven voltage signals and maximum-liklihood fitting \cite{norrelykke_power_2010,dawson_spectral_2019} to Eq.~\eqref{eq:voltage_PSD_air}, we can learn \cite{hillberry_weighing_2020} $\rho=1.7(1)\,{\rm g/cm^3}$, $\kappa=21.3(7)\,{\rm fN/nm}$, and $\beta=2.1(1)\,{\rm mV/ nm}$. At high frequencies, the spectrum \eqref{eq:voltage_PSD_air}
decays as $\sim \omega^{-4}$ until the detector's constant noise floor $\chi = 0.49(2)\,{\rm \upmu V^2 /Hz}$ dominates the signal. Our detector's narrow-band position sensitivity is therefore $\sqrt{\chi}/\beta = \qty{333(21)}{fm/\sqrt{Hz}}$. The inset of Fig.~\ref{fig:calibration} (a) shows that subtle hydrodynamic effects described by Eq.~\eqref{eq:voltage_PSD} are perceptible in thermally driven motion above $\sim \qty{50}{kHz}$, but may be ignored for calibration purposes by restricting the fit domain. In the next section, we will calculate the response of the microsphere to a harmonic acoustic wave.

\subsection{Signal} 
When impinging on the trapped microsphere along the direction $x$ of position measurement, a sound wave of fluid velocity $u$ and acoustic pressure $p$ applies an external force  \cite{temkin_velocity_1976} $F_{\rm ext} = F_{\nabla}(p) + F_{\rm d}(-u)$. The pressure gradient force is $F_{\nabla}(p)=-4 \pi R^3 \nabla p /3$. Using Euler's (linearized) equation $\nabla p = - \rho_{\rm f} \dot{\bf u}$, the pressure gradient force is  $F_{\nabla} = \delta m \dot{u} = 2 \gamma_0 \tau_{\rm f} \dot{u}/9$. Taking $F_{\rm th}=0$, one can solve the equation of motion \eqref{eq:eom} in the frequency domain for the transfer function $H(\omega) = v(\omega) / u(\omega)$, yielding
\begin{equation} \label{eq:transfer}
    H(\omega) = \frac{\gamma(\omega) - i \omega \delta m}{ \gamma(\omega) - i \omega m + i\kappa / \omega}.
\end{equation}
The transfer function, shown in Fig.~\ref{fig:calibration} (b), describes the microsphere's velocity amplitude and phase relative to that of the fluid. Though $\gamma(\omega)\approx \gamma_0$ is appropriate for thermal fluctuations and system calibration in air, driven motion can occur at much higher frequencies, so we retain all three terms in Eq.~\eqref{eq:hydro_damping}.  For example, at 1 MHz, taking $\gamma(\omega)\approx \gamma_0$ underestimates the amplitude of $H$ by a factor of $\sim2$ and overestimates the phase by $\sim \pi/6$ radians. The primary correction to $H$ beyond $\gamma(\omega)\approx \gamma_0$ comes from the history term in Eq.~\eqref{eq:hydro_damping}; the added mass and pressure gradient effects are both proportional to the density ratio $\delta$ and hence small in air. We retain all terms so that our model remains valid for liquid media for which $\delta \sim 0.1 - 1$.

The detector's voltage signal is converted to an acoustic velocity signal using a frequency domain deconvolution $u(t) = \mathcal{F}^{-1}[\mathcal{F}[V(t)] / \psi_u(\omega)]$ where $\mathcal{F}$ is the Fourier transform, and the microsphere's frequency-dependent velocity sensitivity is
\begin{equation} \label{eq:sensitivity}
    \psi_{u}(\omega) = \frac{-i \beta H(\omega)}{\omega}\,.
\end{equation}
The sensitivity is proportional to the transfer function $H$, the calibration factor $\beta$, and the factor $-i/\omega$ that affects the required position-to-velocity derivative. For experimental data sampled at a rate $1/dt$, the derivative factor consistent with a central finite difference in the time-domain is $- i/\omega \to -i dt / \sin(\omega dt)$~\cite{sunaina_calculating_2018}. Acoustic pressure and velocity are related through the impedance $Z(\omega) = p(\omega) / u(\omega)$, hence the pressure sensitivity is $\psi_p = \psi_u/Z$. For plane acoustic waves $Z = \rho_{\rm f} c_0$ is a constant. We will assume planar acoustic waves throughout and use the factor $Z$ to freely convert between pressure and velocity.

\begin{figure}[htbp]
\includegraphics[width=\columnwidth]{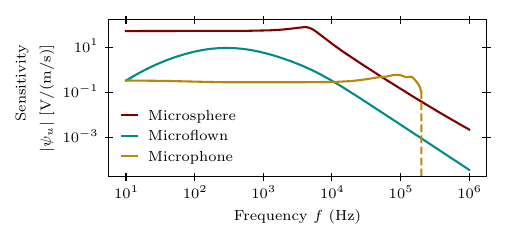} 
\caption{Comparing acoustic detector velocity sensitivities. The microsphere parameters are consistent with the calibration shown in Fig.~\ref{fig:calibration} (a). The microphone sensitivity is provided by the manufacturer and includes corrections for operation without the protective grid and in free-field conditions. The nominal pressure sensitivity is 0.68 mV/Pa and is converted to velocity via the plane-wave impedance of air for comparison with the velocity sensors. The microphone calibration known up to 200 kHz (dashed amber line).
\label{fig:sensitivity} 
    }
\end{figure}

Commercial acoustic detectors are typically calibrated by comparing the sensor's output voltage to a well-characterized input sound under anechoic conditions. By contrast, our thermo-mechanical position calibration and hydrodynamic transfer function enable self-calibration. The sensitivity amplitudes of our commercial microphone and Microflown are provided by the manufacturers and shown in Fig.~\ref{fig:sensitivity} compared to the sensitivity of our microsphere system.

\subsection{Detection limits}
\begin{figure}[htbp]
\includegraphics[width=\columnwidth]{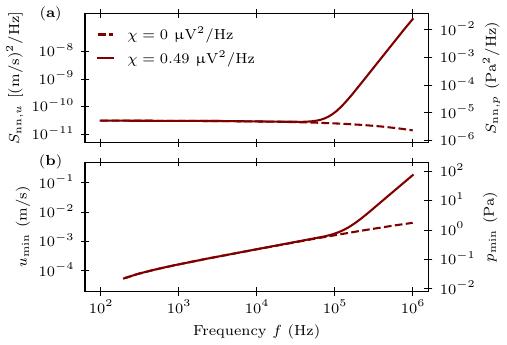}
\caption{
(a) Thermally-driven self-noise spectrum for microsphere-based acoustic sensing.
(b) The minimum-detectable acoustic disturbance estimated from the self-noise spectrum's band-limited variance. In both panels: The solid lines include effects of a constant detection noise floor while the dashed line assumes perfect detection. All other parameters are consistent with the calibration shown in Fig.~\ref{fig:calibration} (a). The left axis quantifies results in terms of acoustic velocity while the right axis converts to pressure via the plane-wave impedance of air.  \label{fig:limits} 
    }
\end{figure}

The above considerations for signal and noise allow us to estimate our microsphere's minimum detectable acoustic signal. A voltage signal derived from only thermal fluctuations \eqref{eq:voltage_PSD} then transformed to a fluid velocity via the sensitivity \eqref{eq:sensitivity} will exhibit a self-noise spectrum [Fig.~\ref{fig:limits} (a)]
\begin{equation}\label{eq:selfnoise}
    %S_{{\rm nn}, u}(\omega) = \frac{S_{VV}(\omega) + \chi}{\lvert s_u \rvert^2}.
    S_{{\rm nn}, u}(\omega) = \frac{S_{VV}(\omega)}{\lvert \psi_u(\omega)\rvert^2} = \frac{4  k_{\rm  B} T {\rm Re}[\gamma(\omega)]}{\lvert \gamma(\omega) - i \omega \delta m\rvert^2}.
\end{equation}
 The self-noise is quite flat and near the DC value $S_{{\rm nn},u}(\omega \to 0) = 4 k_{\rm B} T /\gamma_0$.  
From the self-noise spectrum, the minimum-detectable signal is given by the band-limited variance [Fig.~\ref{fig:limits} (b)]
$
    u_{\rm min} = \sqrt{\int_0^f {\rm d} f' \, S_{{\rm nn}, u}(2 \pi f')}\, .
$
One can include the effects of a constant detector noise floor by making the replacement $S_{VV}(\omega) \to S_{VV}(\omega) + \chi$ in Eq.~\eqref{eq:selfnoise}.

\section{Results} \label{sec:results}
Having established the operating principle and expected performance of optically-trapped microspheres as acoustic sensors, we next describe experimental results. Using a two-channel high-speed digitizer, we record the microsphere signal and either the microphone or the Microflown signal when driven by various sound sources. Each channel is analog-low-pass filtered to 4 MHz then sampled at a rate of $1/dt = \qty{25}{MHz}$ to minimize aliasing. In post processing, the recorded voltage signals are further low-pass filtered by averaging together adjacent points of non-overlapping segments, thereby adjusting the effective sampling rate and signal bandwidth. Once filtered, the voltage signals are converted to either pressure or velocity using the appropriate sensitivity curves.

\subsection{Tone-burst sound source}
\begin{figure}[tbp]
\includegraphics[width=\columnwidth]{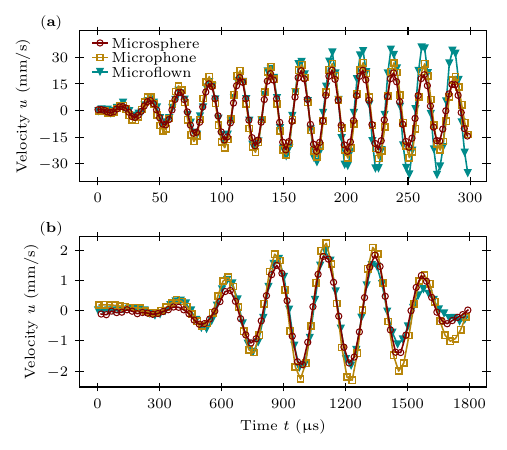}
\caption{Comparing measurements of tone-burst signals between three acoustic sensors. (a) Ten cycles of a 40 kHz tone (9 V peak-to-peak drive voltage). All sensors are post-processed to a bandwidth of 200 kHz (b) Three cycles of a 4 kHz tone (7 V peak-to-peak drive voltage). All sensors are post-processed to a bandwidth of 20 kHz. In both panels, 100 independent trials are averaged, and the origin in time is aligned for each sensor manually. \label{fig:tone_burst} 
    }
\end{figure}

Tone bursts, consisting of a certain number of sinusoidal cycles at a single frequency, provide a simple and repeatable test signal for our various acoustic detectors. In our first set of experiments, we launch tone busts using a function generator to drive piezoelectric buzzers held a distance $\Delta x=\qty{44}{mm}$ from the optical trap. $\Delta x$ is varied by mounting the piezo buzzers on a motorized platform. We drive one of two buzzers at their resonant frequencies $\qty{4}{kHz}$ or $\qty{40}{kHz}$. We observe excellent agreement between our commercially-calibrated reference sensors and our thermo-mechanically calibrated research system, as shown in Fig.~\ref{fig:tone_burst}. The agreement between sensors lessens as source distance $\Delta x$ or time $t$ increases (see Fig.~\ref{fig:tone_burst_pos_scan} of the Appendix). The loss of agreement could be due to a number of effects including acoustic scattering and diffraction, and differences in sensor directivity, placement, and size. 

\subsection{Laser ablation sound source}

A pulsed laser focused to a small spot on a surface can deposit a vast amount of energy in a short amount of time \cite{murray_laser_1999}. This phenomenon has fueled diverse technologies including micro-machining~\cite{liu_laser_1997}, laser-induced-breakdown spectroscopy~\cite{pasquini_laser_2007}, thin film growth~\cite{dijkkamp_preparation_1987}, and a platform for studies of light-plasma interactions~\cite{gibbon_short-pulse_1996}. The sharp acoustic impulse generated by laser ablation has spurred its own research thrusts on non-contact damage detection~\cite{kajiwara_damage_2018}, medical imaging~\cite{wang_photoacoustic_2012}, and scale-modeling of sonic booms~\cite{qin_characteristics_2004}. The impulse has an N-shaped acoustic signature, consisting of a sharp rise, followed by a decay through a zero-crossing into a slower-timescale trough.

In our second set of experiments, we use laser ablation to generate high-frequency-content impulsive sounds to test the high-frequency measurement capabilities of our microsphere-based acoustic sensor. The ablation laser operates at a wavelength of 532 nm with a pulse width of 5 ns and an energy of $\sim7$ mJ. The pulse has a flat-top mode shape that is focused with a 65 mm focal length lens to $\sim \qty{75}{\micro m}$ on an aluminum target. The ablation target, focusing lens, and laser steering mirror are all mounted on the motorized platform used to vary the source distance $\Delta x$. The ablation target is further motorized to rotate and reveal a fresh target spot every ten shots. For this experiment, we do not measure the Microflown signal because of its limited high-frequency sensitivity. 

\begin{figure}[tbp]
\includegraphics[width=\columnwidth]{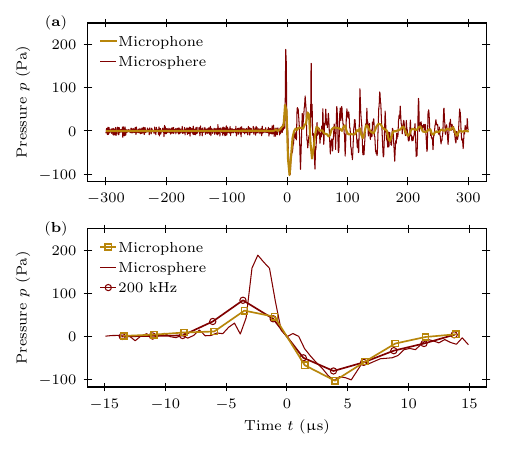}
\caption{Microsphere and microphone response to an acoustic impulse generated by laser ablation, averaged over 10 shots. (a) A trace showing the initial noise level, leading edge arrival, and subsequent reverberations. The microphone is processed with its maximum bandwidth of 200 kHz, and the microsphere is processed with a bandwidth of 1 MHz. The time origin is set to the first zero crossing following the leading edge. (b) A trace of the same impulse over a $20\times$ shorter time window. The solid red line is the microsphere data shown in (a), the open squares are the microphone data, and the open circles are the microsphere data filtered to a bandwidth of 200 kHz. \label{fig:pulsed_laser} 
    }
\end{figure}

Figure \ref{fig:pulsed_laser} shows the microphone and microsphere signals at $\Delta x = \qty{100}{mm}$. It is well known that standard microphones are unable to resolve the rising edge of the acoustic impulse sourced by laser ablation \cite{thomas_acoustical_2017}, necessitating alternative methods such as laser deflection or interference \cite{wissmeyer_looking_2018}. Our results indicate optically-trapped microspheres offer another alternative that is capable of measuring impulsive signals with a $\sim \qty{1}{\micro s}$ rising edge, defined as the time for the signal to change from 10\% to 90\% of its peak value. By comparison, the microphone measures a rise-time of $\sim \qty{5}{\micro s}$. As $\Delta x$ decreases, the microsphere signal becomes more intricate, featuring two or more initial peaks (see Fig.~\ref{fig:pulsed_laser_pos_scan} of the Appendix). The details of these features are very sensitive to the orientation of the target and its lateral offset from the trap center.

\section{Discussion} \label{sec:discussion}
We now turn to a discussion of the results presented in the previous section. We then contextualize the results by reviewing similar work using optically-trapped microspheres for flow measurements. Finally, we outline possible extensions and applications left for future work.

From the tone-burst experiments, we conclude that our microsphere-based acoustic sensor is capable of making calibrated acoustic measurements. All three sensors agree well when converted to the same units, suggesting the plane-wave impedance model is acceptable and that our microsphere calibration and sensing protocol are correct. The laser ablation sound source highlights the microsphere's superior bandwidth in the form of a steeper rising edge and higher peak pressure as compared to the microphone. In the trough portion of the ablation signal, the two sensors are in better quantitative agreement because acoustic variations are slower and therefore less susceptible to band-limited distortion. When the analysis bandwidth of the microsphere is restricted to that of the microphone [open-circles in Fig.~\ref{fig:pulsed_laser} (b)], the rise times and peak pressures are in much better agreement. Unlike the tone-burst sources, shorter source distances $\Delta x$ result in worse agreement between the microsphere and microphone for laser ablation sources. We understand this as a near-field source impedance effect. Indeed, laser ablation acoustic waves are typically modeled as spherical or cylindrical waves for which the impedance is a complex-valued function that approaches to the plane-wave value at large source distances $\Delta x$. Taken together, our experiments show that optically-trapped microspheres enable calibrated and high-bandwidth sensing of an acoustic wave's velocity quadrature. 

Let us next contrast our microsphere-based sensing protocol with other experiments in the recent literature. First, one other work has couched their experiments as acoustic sensing using optically-trapped microspheres~\cite{ohlinger_optically_2012}, but in a dramatically different regime. In that work, a 60 nm gold sphere is trapped in water and imaged at 50 Hz with a camera. Sounds are generated by intensity-modulating a CW laser beam focused onto a nearby cluster of gold nanoparticles at 10 Hz to 50 Hz, or by a needle attached to a 300 Hz loudspeaker. Since the detection method is slow, the methodology hinges on measuring the particle's position variance in response to sound, hence no time-dependent waveforms may be constructed. The authors claim to be able to detect sound power down to a level of -60 ${\rm dB_{re\,1pW}}$. 
Similar frequency-domain analysis of camera-captured microsphere trajectories is used in~\cite{kirchner_direct_2014}, where flow is generated by the rotating flagella bundle of an optically-trapped bacterium, and in~\cite{nedev_microscale_2014} where flow is generated by periodically blocking and unblocking one of two transversly-separated traps, causing a drive particle to periodically jump. In~\cite{bassindale_measurements_2014}, a microsphere is trapped in water contained within a 6.8 MHz, piezo-driven, standing-wave chamber. The time-averaged microsphere position is recorded using a camera at 150 Hz. The steady-state displacement of the microsphere from its equilibrium position maps the standing-wave profile. In a more-recent work termed \emph{optical tweezer-based velocimetry}~\cite{dehnavi_optical_2020}, a position-sensitive detector monitors a microsphere optically trapped in a water-filled sample chamber. The sample chamber is driven at frequencies of 1 Hz - 90 Hz. Velocity amplitudes of \qty{1.5}{\micro m \per s} - \qty{70}{\micro m \per s} are detected in real-time. Such low amplitudes beat the thermal limit by using a Kalman filter to deduce the flow velocity from microsphere position measurements in the presence of Brownian motion. 
In another recent work~\cite{bruot_direct_2021}, a silica microsphere is optically trapped in water and driven transversely at 50 Hz to 400 Hz. An additional 30 smaller polystyrene tracer particles, initially optically trapped at fixed locations near the drive particle, are released upon starting the drive and observed to follow Lissajous trajectories. Compared to previous efforts, our work is unique because it is performed in air, it makes quantitative acoustic field measurements that are bench marked against well-calibrated detectors, and it does so with enough time resolution to observe acoustic waveforms at 4 kHz and 40 kHz, as well as impulsive waveforms with frequency content in the megahertz-range. Like some of the above methods, our method measures the flow velocity of the surrounding fluid. However, instead of inferring flow velocity through microsphere displacement, we rely on microsphere velocity measurements and a hydrodynamic model of the viscous coupling between fluid and microsphere, thereby dramatically increasing the detection bandwidth. 

Our results set up numerous opportunities for follow-up work.  First, incorporating a Kalman filter could increase the signal-to-noise ratio while preserving the ability to self-calibrate. Second, our demonstration was in air, but the theory is equally valid in liquid. Acoustic transduction in a liquid is more efficient than in a gas due to a greater similarity in acoustic impedance between the solid transducer and the medium in which the sound propagates. Therefore, it would be interesting to compare our method to state-of-the-art acoustic sensors for water, such as a needle hydrophone. Finally, since the microsphere measures acoustic velocity, it could be combined with novel opto-acoustic methods that are capable of high-bandwidth pressure measurement to elucidate the impedance of unique sources like blast-waves from laser ablation, surface acoustic waves, and surface vibrations in the near-field. Further, since velocity is a vector-quantity, the microsphere could be useful in sound-source localization, opening the door to several applications. Applications of high-bandwidth acoustic velocity sensing could include locating where a firearm has been discharged, real-time monitoring in proton-therapy for cancer treatment \cite{de-matteis_acoustic_2021,deurvorst_spatial_2022}, and event discrimination in bubble-chamber searches for dark matter \cite{behnke_improved_2011,kozynets_modeling_2019,amole_dark_2019}.

\section{Conclusions} \label{sec:conclusions}
By monitoring an optically-trapped microsphere's instantaneous velocity, we infer fluid flow of sonic, ultrasonic, and impulsive perturbations in air. We validate the accuracy of our technique by comparing tone-burst measurements made with two commercially-available devices, a high-bandwidth pressure microphone and a dual-hot-wire anemometer --- the Microflown --- which measures acoustic velocity. We then test the bandwidth of our sensor by exposing it to impulsive test sounds generated by laser ablation. Beyond the direct extensions mentioned in the previous section, we hope this work inspires other sensing protocols enabled by the resolution of a Brownian particle's instantaneous velocity.

\acknowledgments 
We thank Neal Hall for several useful discussions.

\onecolumngrid
\appendix*
\section{Sound detection results for various source distances}

\begin{figure*}[htbp]
\includegraphics[width=\linewidth]{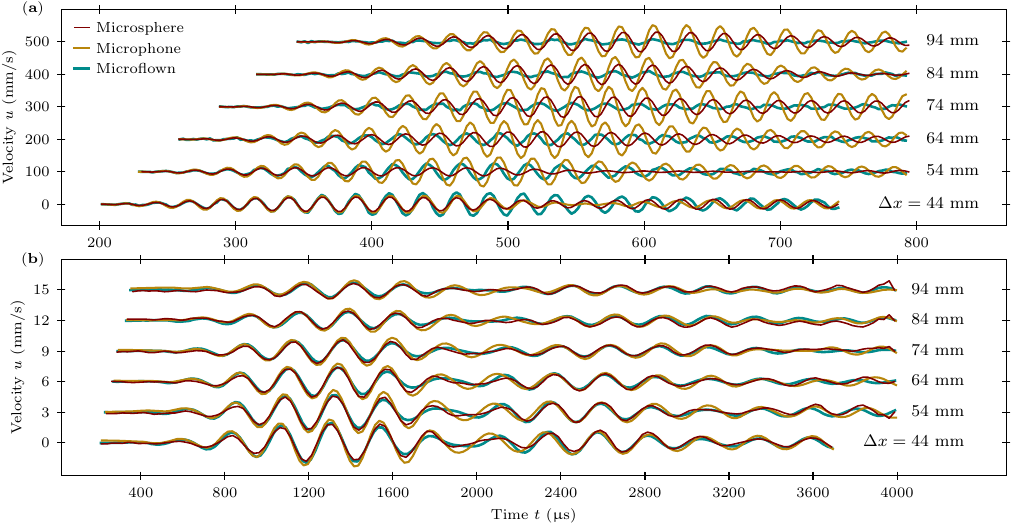}
\caption{Tone burst detection results as in Fig.~\ref{fig:tone_burst}, but for various source distances $\Delta x$ and for (a) 40 kHz and (b) 4 kHz drive. Traces corresponding to different source distances are vertically shifted for clarity.
\label{fig:tone_burst_pos_scan} 
    }
\end{figure*}

\begin{figure*}[htbp]
\includegraphics[width=\linewidth]{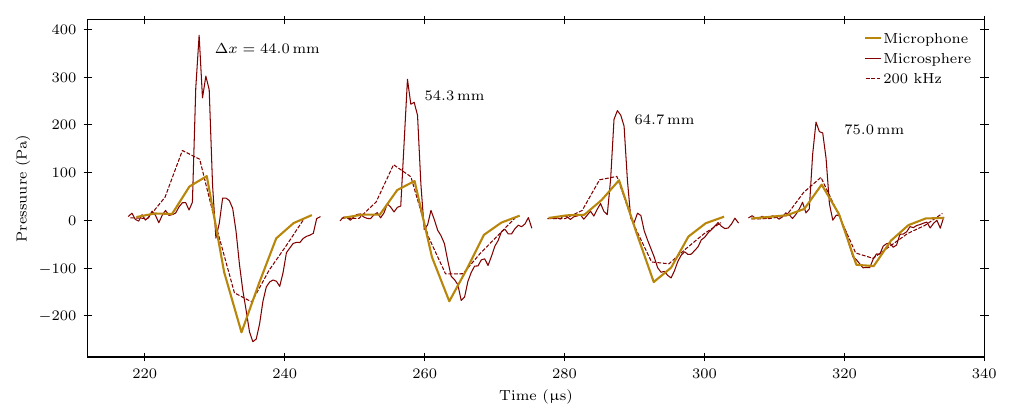}
\caption{ Laser ablation detection results as in Fig.~\ref{fig:pulsed_laser}, but for various source distances $\Delta x$.
\label{fig:pulsed_laser_pos_scan} 
    }
\end{figure*}
\twocolumngrid
~

\clearpage

\bibliography{sound}% Produces the bibliography via BibTeX.

\end{document}